\documentclass[a4paper,12pt]{article}
\usepackage{amsmath}
\usepackage{accents}
\usepackage{bm}
\usepackage[english]{babel}
\usepackage{graphicx}
\usepackage[usenames]{color}
\usepackage{colortbl}
\usepackage[font=footnotesize,figurewithin=none]{caption}
\usepackage{subcaption}
\usepackage{floatrow}
\usepackage[toc,page,title]{appendix}

\begin{document}

\def\Re{\mathop{\rm Re}\nolimits}
\def\Im{\mathop{\rm Im}\nolimits}

\title{Probing the Lee-Yang zeros of a spin bath\\ by correlation functions\\ and entanglement of two spins}

\author{{A. R. Kuzmak$^1$, V. M. Tkachuk$^2$}\\
\medskip
\centerline {\small \it E-Mail: $^1$andrijkuzmak@gmail.com, $^2$voltkachuk@gmail.com}\\
\centerline {\small \it Department for Theoretical Physics, Ivan Franko National University of Lviv,}\\
\centerline {\small \it 12 Drahomanov St., Lviv, UA-79005, Ukraine}}

\maketitle

{\abstract

We study the Lee-Yang zeros of the quantum ferromagnetic Ising bath via the interaction with the two probe spins. Similarly as in paper
[Bo-Bo Wei, Ren-Bao Liu, Phys. Rev. Lett. {\bf 109}, 185701 (2012)] the problem of detecting the zeros is reduced to the exploration of time evolution
of probe spins. As a result, the relation between the Lee-Yang zeros of the bath and correlation functions of the probe system is obtained.
Also we obtain relation between the Lee-Yang zeros and values of the entanglement of probe spins. We apply these results to
the 1D Ising spin model with nearest-neighbor interaction which can be prepared on trapped atoms.

\medskip

}

\section{Introduction \label{sec1}}

In papers \cite{LeeYangZeros1,LeeYangZeros}, Lee and Yang developed a new method based on the analysis of the partition function zeros (Lee-Yang zeros).
The method was used for studying the thermodynamic properties of a general ferromagnetic Ising model defined by the Hamiltonian
\begin{eqnarray}
H_{b}=-\sum_{i,j}J_{ij}^bs_i^{z}s_j^{z}-h\sum_{i}s_i^{z},
\label{bathham}
\end{eqnarray}
where $s_i^{\alpha}$ is the $\alpha=x,y,z$ component of the spin-$1/2$ operator which is determined by the Pauli $\alpha$-matrix as follows $s_i^{\alpha}=\sigma_i^{\alpha}/2$,
$J_{ij}^b\geq 0$ are the interaction couplings and $h$ is the value of the magnetic field. We use the system of units, where the Planck constant
is $\hbar=1$. This means that the energy is measured in the units of frequency. Lee and Yang showed that the zeros of the partition function can be used as a tool for exploring character of phase transitions. The Lee-Yang zeros are
determined by the values of the magnetic field at which the partition function of the system vanishes. They are located in the complex plane
which corresponds to complex values of the magnetic field. Really, the partition function of system (\ref{bathham}) at temperature $T$ can be expressed
as a $2Ns$-th polynomial of $z\equiv\exp{\left(-\beta h\right)}$ as
\begin{eqnarray}
Z\left(\beta, h\right)=\textrm{Tr}\left[e^{-\beta H_{b}}\right]=e^{\beta Nsh}\sum_{n=0}^{2Ns}p_{n}z^{n},
\label{partfunc}
\end{eqnarray}
where $p_n$ is the partition function with zero magnetic field under the constraint that the total
spin of the system has the projection $Ns-n$, $\beta=1/T$ is the inverse temperature and $s$ is the value of each spin in the bath which equals $1/2$.
Here, the Boltzmann constant is taken as unity. So, for certain $z$ partition function (\ref{partfunc}) becomes zero and we denote this zero as $z_n$, where
$n=1,2,\ldots,2Ns$. In \cite{LeeYangZeros} Lee and Yang proved the theorem that the zeros of partition function (\ref{partfunc})
are lying on the unit circle in the complex plane $z$. This fact allows to rewrite the partition function as follows
\begin{eqnarray}
Z\left(\beta, h\right)=p_0e^{\beta Nsh}\prod_{n=1}^{2Ns}\left(z-z_n\right),
\label{partfunc2}
\end{eqnarray}
where $z_n=e^{i\theta_n}$, $\theta_n$ is the polar angle of the $z_n$ zero on the unit circle. This theorem was generalized
to the general ferromagnetic Ising model of an arbitrary high spin \cite{genforarbspin1,genforarbspin2,genforarbspin3} and the other type
of interaction \cite{genforarbint1,genforarbint2,genforarbint3,genforarbint4} including the ferromagnetic anisotropic Heisenberg model \cite{LeeYangZerosahi}.
However, this theorem is not valid to other many-body systems. For example, in the case of antiferromagnetic Ising system the Lee-Yang zeros
are not distributed along the unit circle. It is worth noting that the zeros of the partition function can be considered with respect to  other physical parameters.
For instance, Fisher studied the zeros of the partition function with a complex temperature \cite{FisherZeros}. See also recent paper \cite{zerospartfunbose5}.

Due to the complex nature of the Lee-Yang zeros, it is difficult to implement direct experimental observations of them. The first experimental
research of the density function of zeros on the Lee-Yang circle is based on the analysis of isothermal magnetization of the Ising ferromagnet
FeCl$_2$ in an axial magnetic field \cite{zerospartfuncspin2,ylesdfehfmd}. In papers \cite{zerospartfuncspin3,zerospartfuncspin4}, the study of
the Lee-Yang zeros was considered in the time domain. The authors obtained the relation between the Lee-Yang zeros of the Ising ferromagnet
and the decoherence of the probe spin. This fact allowed to provide direct experimental observation of Lee-Yang zeros
on trimethylphosphite molecule \cite{zerospartfuncspin1}. In \cite{kuzmak2019}, the physical values of the probe spin, which should be measured
to detect the Lee-Yang zeros of an arbitrary high spin Ising ferromagnet, were obtained. Also in paper \cite{zerospartfuncspin7} the connection between
the zeros of the partition function and the two-time correlation function of probe spin was found. The possibility of experimental observation
of Lee-Yang zeros of an interacting Bose gas was proposed in \cite{zerospartfunbose4}. In paper \cite{eddlyz}, it was obtained the method of the detection
of dynamical Lee-Yang zeros. The exploration of the zeros of partition functions of various many body systems can be also found
in papers \cite{zerospartfuncspin5,zerospartfuncspin6,zerospartfunbose1,zerospartfunbose2,zerospartfunbose3,zerospartfunfermi1,zerospartfunfermi2,Xu2019} and references therein.

In previous papers \cite{zerospartfuncspin3,zerospartfuncspin4,zerospartfuncspin1,kuzmak2019} the authors studied the Lee-Yang zeros of
quantum ferromagnetic Ising bath in the context of decoherence of the probe spin. As a result, the connection between the observed values
of the probe spin and Lee-Yang zeros of the bath was found. In the present paper, as a probe system, we consider two interacting spin.
As far as we know, the problem with two probe spins has not been considered in the literature. In this case appear new features
such as an entanglement and correlations between probe spins. These features can be related to the thermodynamic characteristics of the bath. Namely,
exploring the entanglement and correlations as functions of time we can detect the Lee-Yang zeros of the bath. So, we explore the influence of an arbitrary
quantum Ising ferromagnetic bath on the evolution of probe spins. Similarly, as in the case of one probe spin, we obtain the relation between
the partition function of the bath and time-dependence of observed values of the probe system. However, in contrast to the previous case,
where the mean values of one probe spin should be measured, here we propose the correlation functions of two probe spins which should
be measured to detect the Lee-Yang zeros of ferromagnetic bath. Also we investigate the influence of the bath on entanglement of the probe system.
As a result we obtain the connection between the value of entanglement of probe spins and the Lee-Yang zeros.
Such a connection appears for the probe systems that consist of more than one spin. Therefore, this is a completely new way which allows to
detect the Lee-Yang zeros of the ferromagnetic bath via correlations between two probe spins.

The paper is organized in the following way. The relation between the Lee-Yang zeros of the bath and measured values (such as correlation functions)
of probe spins is obtained in Section \ref{sec2}. Also in Section \ref{sec3} we obtain the connection between the Lee-Yang zeros
of the bath and values of the entanglement of probe spins. We present our results for 1D Ising spin ring with nearest-neighbor interaction
which can be prepared on trapped atoms (Section \ref{sec4}). Discussions and conclusions are presented in Section \ref{sec5}.

\section{Connection between the Lee-Yang zeros of the spin bath and observed values of probe spins \label{sec2}}

We consider the general quantum ferromagnetic Ising spin bath defined by Hamiltonian (\ref{bathham}) which interacts with two probe spins $1/2$ described
by an anisotropic Heisenberg model. The interaction between these subsystems is defined by the Ising Hamiltonian.
Also the system is placed in an external magnetic field which is directed along the $z$-axis. The Hamiltonian of the system has the form
\begin{eqnarray}
H=H_{b}+H_{AB}+H_{i},
\label{generalham}
\end{eqnarray}
where
\begin{eqnarray}
H_{AB}=J_{xx}\left(s_A^x s_B^x+s_A^y s_B^y\right)+J_{zz}s_A^z s_B^z+h_0\left(s^z_A +s^z_B\right)
\label{probeham}
\end{eqnarray}
describes the interaction between the probe spins $A$ and $B$ with coupling constants $J_{xx}$, $J_{zz}$, and the interaction of the probe spins with
magnetic field of the value $h_0$, and
\begin{eqnarray}
H_{i}=\lambda \left(s^z_A+s^z_B\right)\sum_{i}s_i^{z}
\label{intham}
\end{eqnarray}
describes the interaction between the bath and probe spins with the coupling constant $\lambda$. Note that instead of the Ising bath the
isotropic Heisenberg bath with interaction $\left(-\sum_{i,j}J_{ij}{\bf s}_i{\bf s}_j\right)$ or some other interactions that commute with total
spin of the system can be used.

So, assuming the spin bath being in thermodynamic equilibrium we study the evolution of system (\ref{generalham}) having started from the state
\begin{eqnarray}
\rho(0)=\vert\Psi(0)\rangle\langle\Psi(0)\vert\frac{e^{-\beta H_{b}}}{Z\left(\beta, h\right)},
\label{initstate}
\end{eqnarray}
where $\vert\psi(0)\rangle=\sum_{m_A,m_B=-1/2}^{1/2}a_{m_A,m_B}\vert m_A,m_B\rangle$ is the initial state of two probe spins $A$ and $B$. Here $\vert m_{A}, m_{B}\rangle$
are the states which define the projections of $A$ and $B$ spins on the $z$-axis with $m_{A}$ and $m_{B}$ eigenvalues, respectively, and $a_{m_A,m_B}$ some complex parameters,
which determine the initial state and satisfy the normalization condition $\sum_{m_A,m_B=-1/2}^{1/2}\vert a_{m_A,m_B}\vert^2=1$. Using the
fact that $H_{b}$, $H_{AB}$ and $H_{i}$ mutually commute this evolution can be expressed as follows
\begin{eqnarray}
\rho(t)=e^{-iHt}\rho(0)e^{iHt}=e^{-iH_{AB}t}e^{-iH_{i}t}\rho(0)e^{iH_{i}t}e^{iH_{AB}t}.\nonumber
\end{eqnarray}
Then the density matrix of the system takes the form
\begin{eqnarray}
&&\rho(t)=\sum_{m_A,m_B=-1/2}^{1/2}\sum_{k_A,k_B=-1/2}^{1/2}a_{m_A,m_B}(t)a_{k_A,k_B}^*(t)\nonumber\\
&&\times e^{-i\lambda\left(m_A-k_A+m_B-k_B\right)t\sum_is_i^z}\frac{e^{-\beta H_{b}}}{Z\left(\beta, h\right)} \vert m_A,m_B\rangle\langle k_A,k_B\vert,
\label{evolution}
\end{eqnarray}
where
\begin{eqnarray}
&&a_{\frac{1}{2},\frac{1}{2}}(t)=a_{\frac{1}{2},\frac{1}{2}}e^{-ih_0t}e^{-i\frac{J_{zz}t}{4}},\quad  a_{-\frac{1}{2},-\frac{1}{2}}(t)=a_{-\frac{1}{2},-\frac{1}{2}}e^{ih_0t}e^{-i\frac{J_{zz}t}{4}},\nonumber\\
&&a_{\frac{1}{2},-\frac{1}{2}}(t)=e^{i\frac{J_{zz}t}{4}}\left(a_{\frac{1}{2},-\frac{1}{2}}\cos\left(\frac{J_{xx}t}{2}\right)-ia_{-\frac{1}{2},\frac{1}{2}}\sin\left(\frac{J_{xx}t}{2}\right)\right),\nonumber\\
&&a_{-\frac{1}{2},\frac{1}{2}}(t)=e^{i\frac{J_{zz}t}{4}}\left(a_{-\frac{1}{2},\frac{1}{2}}\cos\left(\frac{J_{xx}t}{2}\right)-ia_{\frac{1}{2},-\frac{1}{2}}\sin\left(\frac{J_{xx}t}{2}\right)\right).\nonumber
\end{eqnarray}
The evolution of the $A$ and $B$ spins can be obtained if in expression (\ref{evolution}) the average over the states of the remaining system is done.
Then the reduced density matrix for probe spins reads
\begin{eqnarray}
&&\rho_{AB}(t)=\sum_{m_A,m_B=-1/2}^{1/2}\sum_{k_A,k_B=-1/2}^{1/2}a_{m_A,m_B}(t)a_{k_A,k_B}^*(t)\nonumber\\
&&\times\frac{\textrm{Tr}\left[ e^{-\beta H_{b}-i\lambda\left(m_A-k_A+m_B-k_B\right)t\sum_is_i^z}\right]}{Z\left(\beta, h\right)} \vert m_A,m_B\rangle\langle k_A,k_B\vert,
\label{probeevol}
\end{eqnarray}
where $\textrm{Tr} [\ldots]$ is calculated over the bath system.
This expression is represented by the partition functions of the bath with different complex magnetic fields $h-i\lambda(m_A-k_A+m_B-k_B)t/\beta$
as follows
\begin{eqnarray}
&&\rho_{AB}(t)=\sum_{m_A,m_B=-1/2}^{1/2}\sum_{k_A,k_B=-1/2}^{1/2}a_{m_A,m_B}(t)a_{k_A,k_B}^*(t)\nonumber\\
&&\times\frac{Z\left(\beta, h-i\lambda(m_A-k_A+m_B-k_B)t/\beta\right)}{Z\left(\beta, h\right)} \vert m_A,m_B\rangle\langle k_A,k_B\vert.
\label{probeevolpf}
\end{eqnarray}
Depending on the values of $m_{A(B)}$ and $k_{A(B)}$, this expression contains two types of the partition functions with
$h\pm i\lambda t/\beta$ and $h\pm i2\lambda t/\beta$ magnetic fields. Each of these functions vanish when the evolution time $t$
is such that $\exp{\left(-\beta h+i\lambda (m_A-k_A+m_B-k_B)t\right)}$ equals to Lee-Yang zeros. To obtain these moments of time
we rewrite these partition functions using representation (\ref{partfunc2}). Then, we get
\begin{eqnarray}
&&\frac{Z\left(\beta, h-i\lambda(m_A-k_A+m_B-k_B)t/\beta\right)}{Z\left(\beta, h\right)}\nonumber\\
&&=\frac{e^{-iNs\lambda(m_A-k_A+m_B-k_B)t}\prod_{n=1}^{2Ns}\left(e^{-\beta h+i\lambda(m_A-k_A+m_B-k_B)t}-e^{i\theta_n}\right)}{\prod_{n=1}^{2Ns}\left(e^{-\beta h}-e^{i\theta_n}\right)}.
\label{probeevolpf1}
\end{eqnarray}
In the case of $h=0$, the Lee-Yang zeros of the above partition function lie on the time axis. Indeed, for the zero magnetic field the partition function
(\ref{probeevolpf1}) takes the form
\begin{eqnarray}
&&\frac{Z\left(\beta, -i\lambda(m_A-k_A+m_B-k_B)t/\beta\right)}{Z\left(\beta,h=0\right)}\nonumber\\
&&=\frac{e^{-iNs\lambda(m_A-k_A+m_B-k_B)t}\prod_{n=1}^{2Ns}\left(e^{i\lambda(m_A-k_A+m_B-k_B)t}-e^{i\theta_n}\right)}{\prod_{n=1}^{2Ns}\left(1-e^{i\theta_n}\right)}.
\label{probeevolpf2}
\end{eqnarray}
It is easy to verify that the Lee-Yang zeros of the ferromagnetic model are lying symmetrically with respect to the real axis. This fact follows from the
real nature of $p_n$ in partition function (\ref{partfunc}). Taking into account this property we represent partition function (\ref{probeevolpf2})
in the form
\begin{eqnarray}
&&\frac{Z\left(\beta, -i\lambda(m_A-k_A+m_B-k_B)t/\beta\right)}{Z\left(\beta,h=0\right)}\nonumber\\
&&=\frac{\prod_{n=1}^{2Ns}\left(\cos\left(\frac{\lambda(m_A-k_A+m_B-k_B)t}{2}\right)-\cos\left(\theta_n-\frac{\lambda(m_A-k_A+m_B-k_B)t}{2}\right)\right)}{\prod_{n=1}^{2Ns}\left(1-\cos\theta_n\right)}.
\label{probeevolpf3}
\end{eqnarray}
This expression vanishes when $\lambda(m_A-k_A+m_B-k_B)t=\theta_n$. So, we obtain the values of times
$t_n=\theta_n/\left(\lambda(m_A-k_A+m_B-k_B)\right)$ which correspond to the Lee-Yang zeros of bath.
Also, we can see that for the partition function with magnetic fields $\pm i2\lambda t/\beta$
the Lee-Yang zeros are achieved twice faster than for the partition function with magnetic fields $\pm i\lambda t/\beta$.
So, the problem of observing the Lee-Yang zeros of the bath is reduced to the study of time evolution of the probe spins.
For this purpose, we propose to measure the correlation functions of the probe spins.
It is easy to verify that the $\langle s_A^xs_B^z\rangle$ and $\langle s_A^ys_B^z\rangle$ functions are expressed by the partition function of bath system
as follows
\begin{eqnarray}
\langle s_A^xs_B^z\rangle &=& \textrm{Tr}\left[\rho_{AB}(t)s_A^xs_B^z\right]=\frac{Z\left(\beta,-i\lambda t/\beta\right)}{Z\left(\beta, h=0\right)}\nonumber\\
&&\times \frac{1}{2}\Re\left(a_{\frac{1}{2},\frac{1}{2}}^*(t)a_{-\frac{1}{2},\frac{1}{2}}(t)-a_{\frac{1}{2},-\frac{1}{2}}(t)a_{-\frac{1}{2},-\frac{1}{2}}^*(t)\right)
\label{averagevalue1}
\end{eqnarray}
and 
\begin{eqnarray}
\langle s_A^ys_B^z\rangle &=& \textrm{Tr}\left[\rho_{AB}(t)s_A^ys_B^z\right]=\frac{Z\left(\beta,-i\lambda t/\beta\right)}{Z\left(\beta, h=0\right)}\nonumber\\
&&\times \frac{1}{2}\Im\left(a_{\frac{1}{2},\frac{1}{2}}^*(t)a_{-\frac{1}{2},\frac{1}{2}}(t)+a_{\frac{1}{2},-\frac{1}{2}}(t)a_{-\frac{1}{2},-\frac{1}{2}}^*(t)\right),
\label{averagevalue2}
\end{eqnarray}
respectively. As we can see that these mean values allow to study the time behavior of the partition function with magnetic field $-i\lambda t/\beta$.
It is important to note that similar results we obtain for mean values of $s_A^zs_B^x$ and $s_A^zs_B^y$ operators. To study the time behavior of the partition function
with magnetic field $-i2\lambda t/\beta$ the mean values of the operators $s_A^xs_B^x$, $s_A^ys_B^y$ should be measured and then
the following difference constructed
\begin{eqnarray}
\langle s_A^xs_B^x\rangle-\langle s_A^ys_B^y\rangle=\frac{Z\left(\beta,-i2\lambda t/\beta\right)}{Z\left(\beta, h=0\right)}\Re \left(a_{\frac{1}{2},\frac{1}{2}}^*(t)a_{-\frac{1}{2},-\frac{1}{2}}(t)\right).
\label{averagevalue3}
\end{eqnarray}
Also we can study the time behavior of the partition function
with magnetic field $-i2\lambda t/\beta$ using the $\langle s_A^xs_B^y\rangle+\langle s_A^ys_B^x\rangle$ correlation function. The relation which appears here is
the following
\begin{eqnarray}
\langle s_A^xs_B^y\rangle+\langle s_A^ys_B^x\rangle=\frac{Z\left(\beta,-i2\lambda t/\beta\right)}{Z\left(\beta, h=0\right)}\Im \left(a_{\frac{1}{2},\frac{1}{2}}^*(t)a_{-\frac{1}{2},-\frac{1}{2}}(t)\right).
\label{averagevalue4}
\end{eqnarray}
So, to detect the Lee-Yang zeros of the bath defined by Hamiltonian (\ref{bathham}) the correlation functions
of the probe spins as functions of time should be measured. Note that connections similar to (\ref{averagevalue1}), (\ref{averagevalue2}), (\ref{averagevalue3}) and (\ref{averagevalue4}) exist in the case of
isotropic Heisenberg interaction between the spins of bath. However, in this case the partition function is defined by the ferromagnetic Heisenberg model.
Also it should be noted that we do not propose to measure the
following correlation functions $\langle s_A^zs_B^z\rangle$, $\langle s_A^xs_B^x\rangle+\langle s_A^ys_B^y\rangle$ and
$\langle s_A^xs_B^y\rangle-\langle s_A^ys_B^x\rangle$ because they do not depend on the partition function of the bath system. These
functions depend only on the initial state and the interaction coupling of the probe spins. They have the following form
\begin{eqnarray}
&&\langle s_A^zs_B^z\rangle= \frac{1}{4}\left(\vert a_{\frac{1}{2},\frac{1}{2}}\vert^2 - \vert a_{\frac{1}{2},-\frac{1}{2}}(t)\vert^2- \vert a_{-\frac{1}{2},\frac{1}{2}}(t)\vert^2 + \vert a_{-\frac{1}{2},-\frac{1}{2}}\vert^2\right),\nonumber\\
&&\langle s_A^xs_B^x\rangle+\langle s_A^ys_B^y\rangle=\Re \left(a_{\frac{1}{2},-\frac{1}{2}}(t)a_{-\frac{1}{2},\frac{1}{2}}^*(t)\right), \nonumber\\
&&\langle s_A^xs_B^y\rangle-\langle s_A^ys_B^x\rangle=\Im \left(a_{\frac{1}{2},-\frac{1}{2}}(t)a_{-\frac{1}{2},\frac{1}{2}}^*(t)\right).
\label{averagevalue5}
\end{eqnarray}

Finally, it is worth noting that the Lee-Yang zeros of the bath can be detected by the single-spin correlation functions $\langle s_A^x+s_B^x\rangle$
and $\langle s_A^y+s_B^y\rangle$. These functions are defined by the partition function of the bath in the following way
\begin{eqnarray}
&&\langle s_A^x+s_B^x\rangle=\frac{Z\left(\beta,-i\lambda t/\beta\right)}{Z\left(\beta, h=0\right)}\Re \left[\left(a_{\frac{1}{2},\frac{1}{2}}(t)+a_{-\frac{1}{2},-\frac{1}{2}}(t)\right)\left(a_{\frac{1}{2},-\frac{1}{2}}^*(t)+a_{-\frac{1}{2},\frac{1}{2}}^*(t)\right)\right],\nonumber\\
&&\langle s_A^y+s_B^y\rangle=-\frac{Z\left(\beta,-i\lambda t/\beta\right)}{Z\left(\beta, h=0\right)}\Im \left[\left(a_{\frac{1}{2},\frac{1}{2}}(t)-a_{-\frac{1}{2},-\frac{1}{2}}(t)\right)\left(a_{\frac{1}{2},-\frac{1}{2}}^*(t)+a_{-\frac{1}{2},\frac{1}{2}}^*(t)\right)\right].\nonumber\\
\label{averagevalue6}
\end{eqnarray}
For $\langle s_A^z+s_B^z\rangle$ correlation function we obtain a similar dependence on the initial state and interaction coupling of the probe spins
as in case of the $\langle s_A^zs_B^z\rangle$ correlation function (\ref{averagevalue5}). Note that in the case of non-interacting probe spins
expressions (\ref{averagevalue6}) up to a factor takes the form as in the case of one probe spin-$1$ obtained in paper \cite{kuzmak2019}.

\section{Connection between the Lee-Yang zeros of the spin bath and entanglement of probe spins \label{sec3}}

It is interesting to study the influence of the bath on the entanglement of probe spins. This knowledge allows us to find the connection
between the Lee-Yang zeros of the bath and the values of entanglement of the probe spins. So, in this section we find these connections.
To determine the value of entanglement of the probe system, we use the Wootters definition
of concurrence \cite{wootters1998}
\begin{eqnarray}
C(\rho)=\max\{0,\omega_1-\omega_2-\omega_3-\omega_4\},
\label{wootters}
\end{eqnarray}
where $\omega_i$ are the eigenvalues, in decreasing order, of the Hermitian matrix $R=\sqrt{\sqrt{\rho}\tilde{\rho}\sqrt{\rho}}$. Here,
$\tilde{\rho}=16 s_A^y s_B^y \rho^*s_A^y s_B^y$. Note that $\omega_i$ are real and positive numbers.
For calculations it is convenient to use the eigenvalues of the non-Hermitian matrix $\rho\tilde{\rho}$ which have the form $\omega_i^2$.
Let us calculate the concurrence of $A$ and $B$ spins for the moment of time $t_n$ that corresponds to zero $z_n$ of partition function
$Z\left(\beta,-i\lambda t/\beta\right)$. Then, the eigenvalues of matrix $R$ constructed by density matrix (\ref{probeevolpf})
in the case of $h=0$ have the form
\begin{eqnarray}
&&\omega_1=2\left\vert a_{\frac{1}{2},-\frac{1}{2}}(t_n) a_{-\frac{1}{2},\frac{1}{2}}(t_n)\right\vert,\quad \omega_2=0,\nonumber\\
&&\omega_{3,4}=\left\vert a_{\frac{1}{2},\frac{1}{2}} a_{-\frac{1}{2},-\frac{1}{2}}\right\vert\left(1\pm\frac{Z\left(\beta,-i2\lambda t_n/\beta\right)}{Z\left(\beta, h=0\right)}\right).
\label{woottersAB}
\end{eqnarray}
The detailed derivation of these eigenvalues is presented in Appendix \ref{appa}. Depending on the ratio between $\omega_i$ we obtain different results
for concurrence. In the case of $\omega_1\geq\omega_{3,4}$ the concurrence has the form
\begin{eqnarray}
C\left({\rho_{AB}(t_n)}\right)=2\left(\left\vert a_{\frac{1}{2},-\frac{1}{2}}(t_n) a_{-\frac{1}{2},\frac{1}{2}}(t_n)\right\vert-\left\vert a_{\frac{1}{2},\frac{1}{2}} a_{-\frac{1}{2},-\frac{1}{2}}\right\vert\right).
\label{concomega1}
\end{eqnarray}
Otherwise, if $\omega_{3,4}>\omega_1$, the concurrence reduces to the form
\begin{eqnarray}
C\left({\rho_{AB}(t_n)}\right)=2\left( \left\vert a_{\frac{1}{2},\frac{1}{2}} a_{-\frac{1}{2},-\frac{1}{2}}\frac{Z\left(\beta,-i2\lambda t_n/\beta\right)}{Z\left(\beta, h=0\right)}\right\vert -\left\vert a_{\frac{1}{2},-\frac{1}{2}}(t_n) a_{-\frac{1}{2},\frac{1}{2}}(t_n)\right\vert \right).
\label{concomega2}
\end{eqnarray}

At other moments of time, which do not correspond to the Lee-Yang zeros of the partition function
$Z\left(\beta,-i\lambda t/\beta\right)$, there is no simple equation describing the time-dependence behavior of the concurrence of the probe spins defined by
density matrix (\ref{probeevolpf}). However, we can study the concurrence of this system in the cases when the evolution happens on the subspaces
spanned by the basis vectors: 1. $\vert \frac{1}{2},-\frac{1}{2}\rangle$, $\vert -\frac{1}{2},\frac{1}{2}\rangle$
($a_{\frac{1}{2},\frac{1}{2}}$=$a_{-\frac{1}{2},-\frac{1}{2}}$=0) and
2. $\vert \frac{1}{2},\frac{1}{2}\rangle$, $\vert -\frac{1}{2},-\frac{1}{2}\rangle$ ($a_{\frac{1}{2},-\frac{1}{2}}$=$a_{-\frac{1}{2},\frac{1}{2}}$=0), respectively.
In the first case the bath cannot impact the evolution of the probe spins and as a result the concurrence depends only on the interaction coupling between spins
as follows
\begin{eqnarray}
C\left(\rho_{AB}(t)\right)=2\left\vert a_{\frac{1}{2},-\frac{1}{2}}(t) a_{-\frac{1}{2},\frac{1}{2}}(t)\right\vert.
\label{woottersAB1}
\end{eqnarray}
Here depending on the initial state the concurrence varies in the range $C\in[0,1]$.
In the second case we obtain the influence of the bath on the concurrence of the probe system
\begin{eqnarray}
C\left(\rho_{AB}(t)\right)=2\left\vert a_{\frac{1}{2},\frac{1}{2}} a_{-\frac{1}{2},-\frac{1}{2}}\frac{Z\left(\beta,-i2\lambda t/\beta\right)}{Z\left(\beta, h=0\right)}\right\vert.
\label{woottersAB2}
\end{eqnarray}
Here the maximal value of entanglement is determined by the initial state. As we can see, the concurrence varies in the range
$C\in\left[0,2\left\vert a_{\frac{1}{2},\frac{1}{2}} a_{-\frac{1}{2},-\frac{1}{2}}\right\vert\right]$.
Here the zeros of the partition function $Z\left(\beta,-i2\lambda t/\beta\right)$ coincide with the zeros of concurrence.

In these cases for the study of entanglement the geometric measure
\begin{eqnarray}
E(\rho_{AB}(t))=\frac{1}{2}\left(1-\sqrt{1-16\langle s_A^x s_B^x\rangle^2-16\langle s_A^y s_B^x\rangle^2}\right)
\label{geommeasure}
\end{eqnarray}
obtained in paper \cite{frydryszak2017} can be applied. Here $\langle s_A^x s_B^x\rangle$ and $\langle s_A^y s_B^x\rangle$ are the correlation functions
calculated with density matrix (\ref{probeevolpf}) in the case of $h=0$. Then the concurrences in equations (\ref{woottersAB1}) and (\ref{woottersAB2})
are defined by these correlations functions as follows
\begin{eqnarray}
C(\rho_{AB}(t))=4\sqrt{\langle s_A^x s_B^x\rangle^2+\langle s_A^y s_B^x\rangle^2}.
\label{relconcorr}
\end{eqnarray}
This fact allows to measure it on an experiment.

Let us consider the results obtained in Sections \ref{sec2} and \ref{sec3} on the real physical system.

\section{Application to the spin ring \label{sec4}}

\begin{figure}[!!h]
\center{\includegraphics[scale=0.5, angle=0.0, clip]{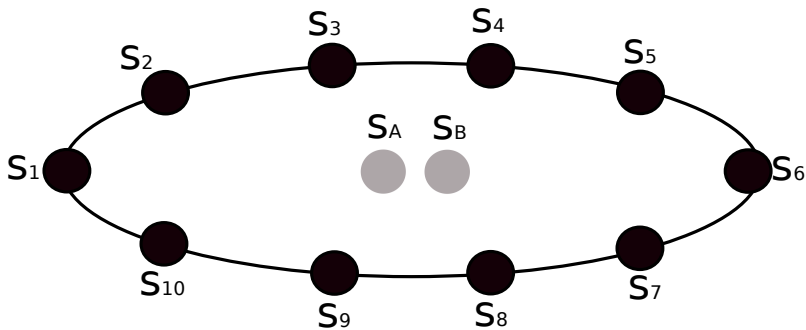}}
\caption{Spin ring. The model consists of ten spins $s_i$ (black circles) as the bath and two probe spin $s_A$ and $s_B$ (gray circle).}
\label{spinring}
\end{figure}

In this section we apply our results to the ring of $N$ spins as the bath and two probe spins placed in the center of this ring (Fig.~\ref{spinring}).
Each of the probe spin interacts with each spins of the bath with the same interaction coupling $\lambda$.
We assume that the interaction between bath spins is described by the 1D ferromagnetic Ising model with nearest-neighbor interaction $J^b$,
and the interaction between the probe spins is isotropic and is defined by the coupling constant $J$. Note that the effective spin ring
can be prepared on trapped ions \cite{SchrodCat1,EQSSTI,QSDEGHTI} or ultracold atoms
\cite{opticallattice1,opticallattice2,opticallattice4}. In paper \cite{sgssrwcsna}, it was presented the method for simulation of the ground
states of spin rings with cavity-assisted neutral atoms.

\begin{figure}[]
\includegraphics[scale=0.34, angle=0.0, clip]{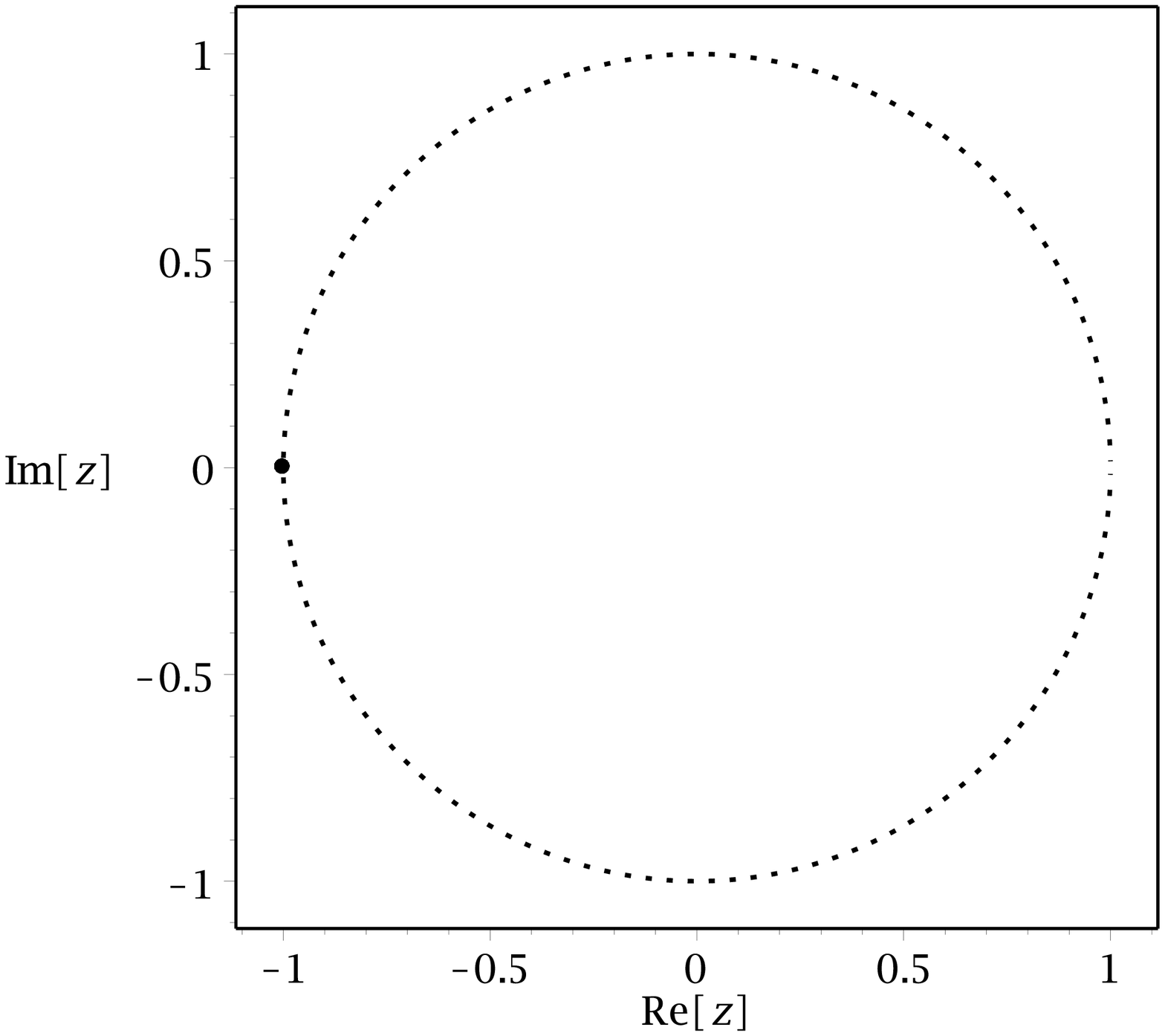}
\includegraphics[scale=0.34, angle=0.0, clip]{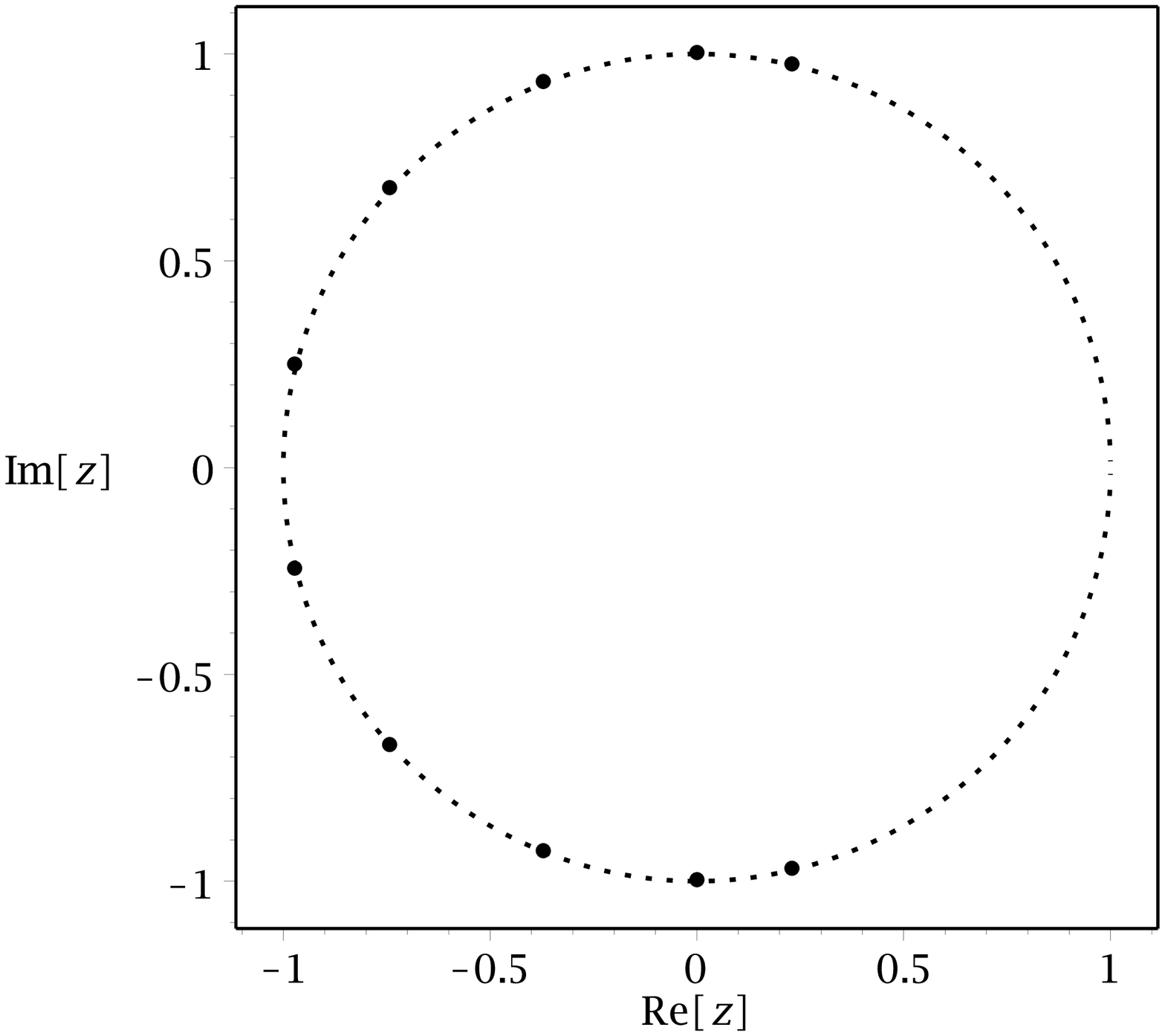}\\
\subcaptionbox{\label{}}{\includegraphics[scale=0.31, angle=0.0, clip]{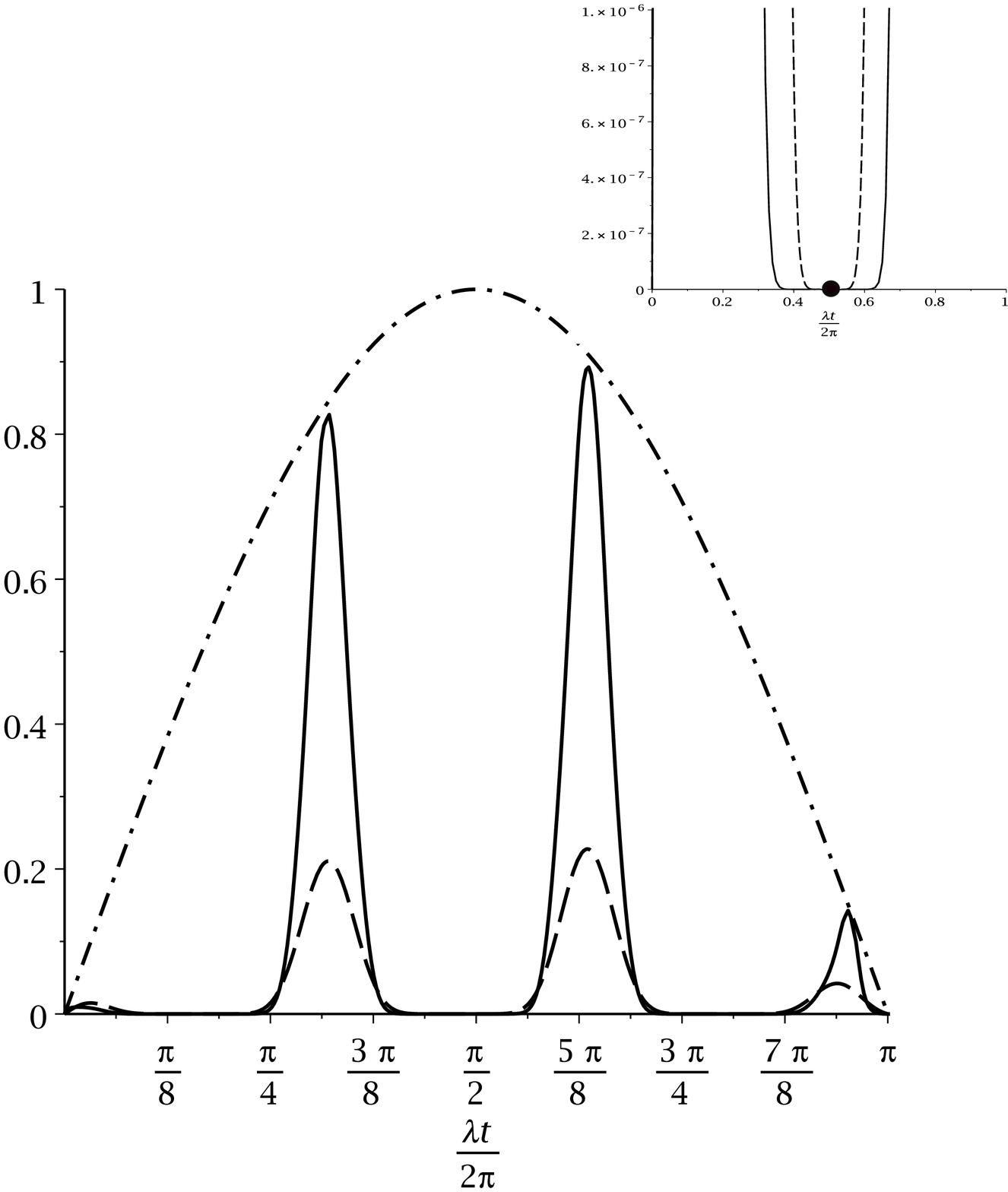}}
\subcaptionbox{\label{}}{\includegraphics[scale=0.31, angle=0.0, clip]{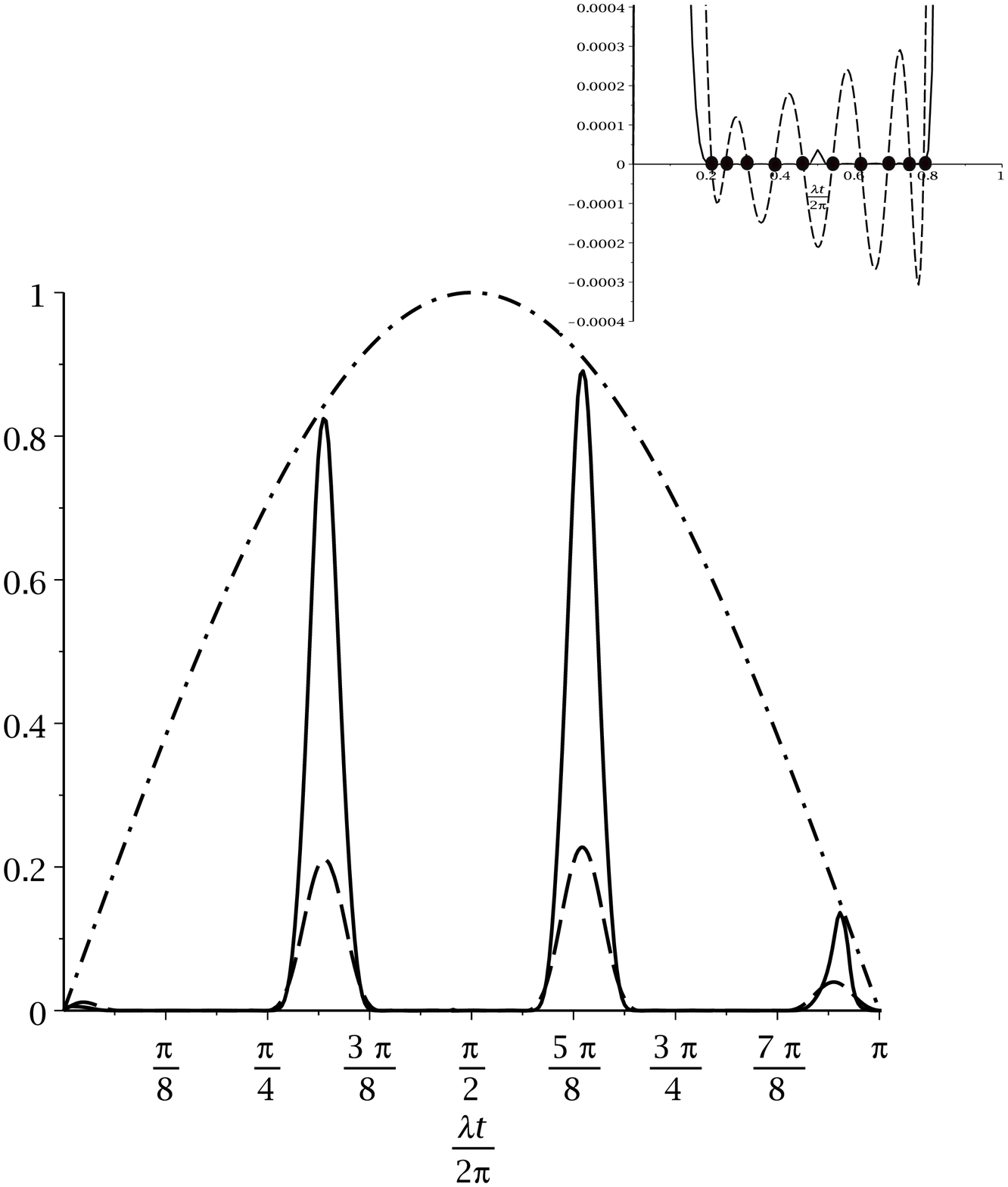}}
\caption{The connection between the Lee-Yang zeros of ten spins $1/2$ Ising ferromagnet with nearest-neighbor interaction (upper figures) and
concurrence of probe spins (solid curve), mean value of the probe spins operator $s_A^ys_B^z$
(\ref{connspinring}) (dashed curve) (lower figures). The results are presented for different temperatures: (a) $T=\infty$ and (b) $T=J^b$.
The upper figures show the positions of the Lee-Yang zeros of bath in the complex plane. The lower figures display the positions
of the Lee-Yang zeros in the time domain. In the insets, the moments of time when the Lee-Yang zeros are reached are marked by the solid circles.
The dash-dotted curve shows behavior of concurrence of the probe spins without the influence of the bath. As we can see, for high temperature only one Lee-Yang zeros exists.
This is a so-called degenerate case. However, the degeneration begins to disappear with decreasing temperature. This fact is clearly visible both
in the complex plane and in the time domain. The interaction between the probe spins is
isotropic and is defined by the coupling constant $J=\lambda/2\pi$.}
\label{lyzspinring1}
\end{figure}

\begin{figure}[]
\includegraphics[scale=0.34, angle=0.0, clip]{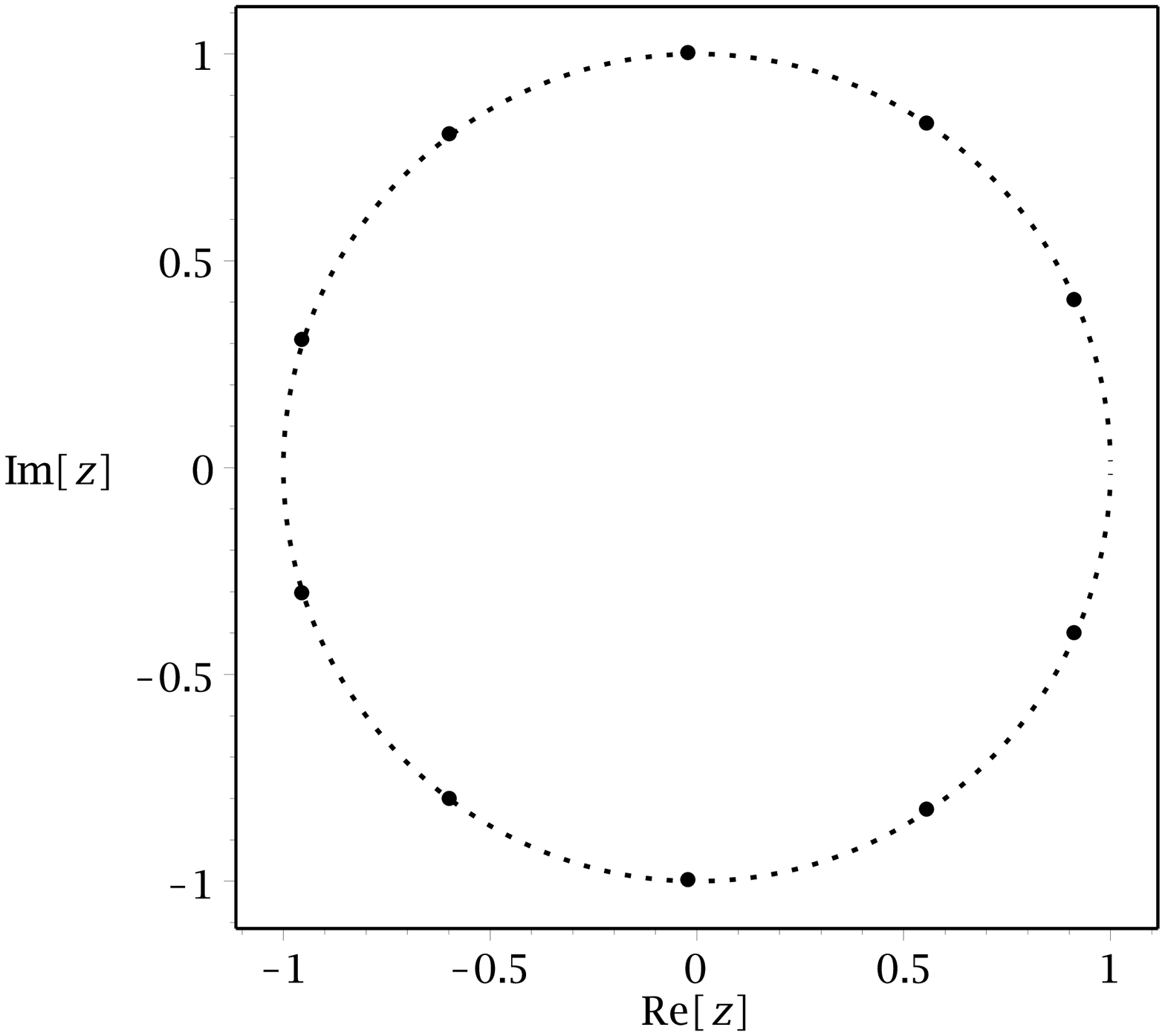}
\includegraphics[scale=0.34, angle=0.0, clip]{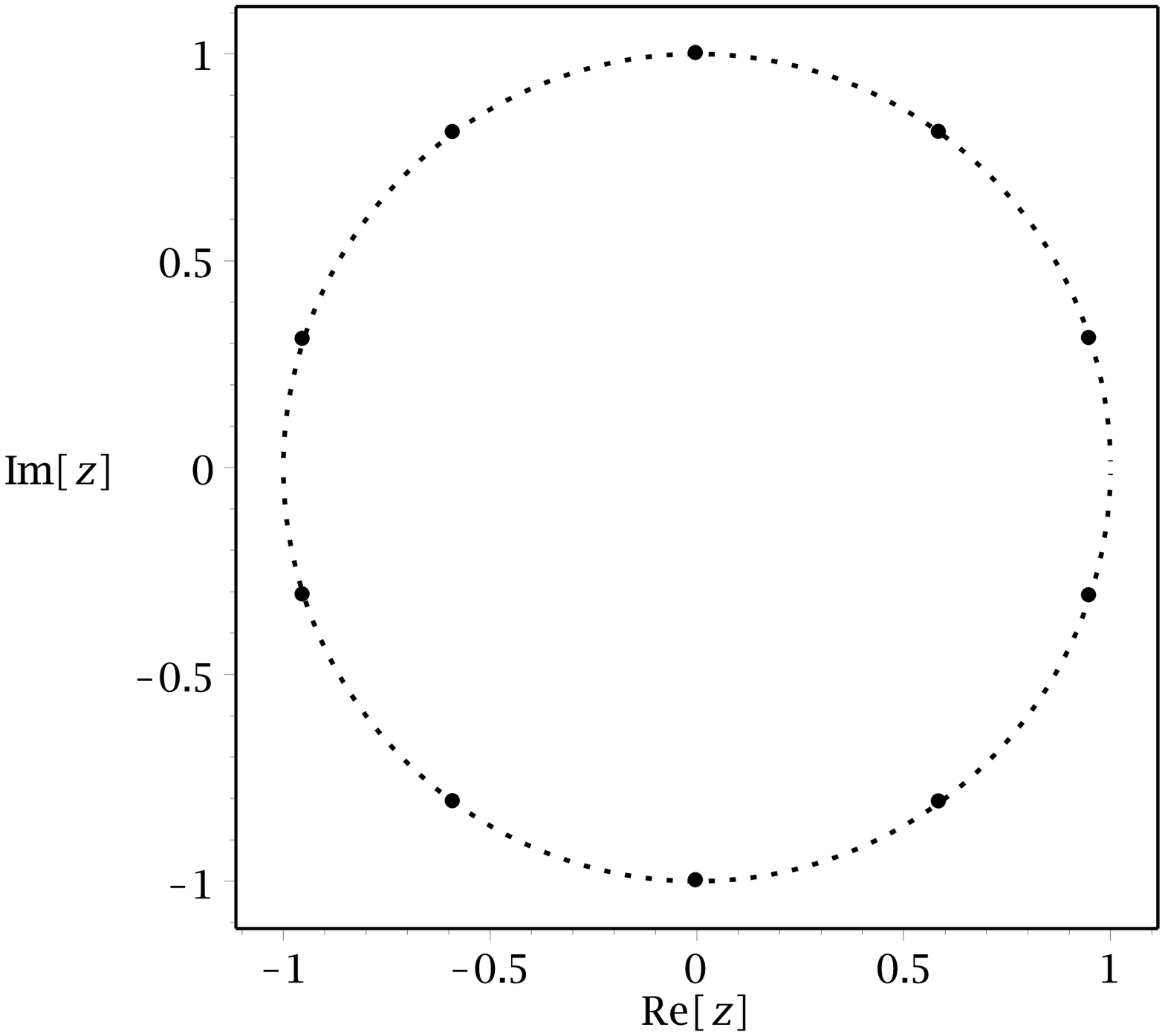}\\
\subcaptionbox{\label{}}{\includegraphics[scale=0.31, angle=0.0, clip]{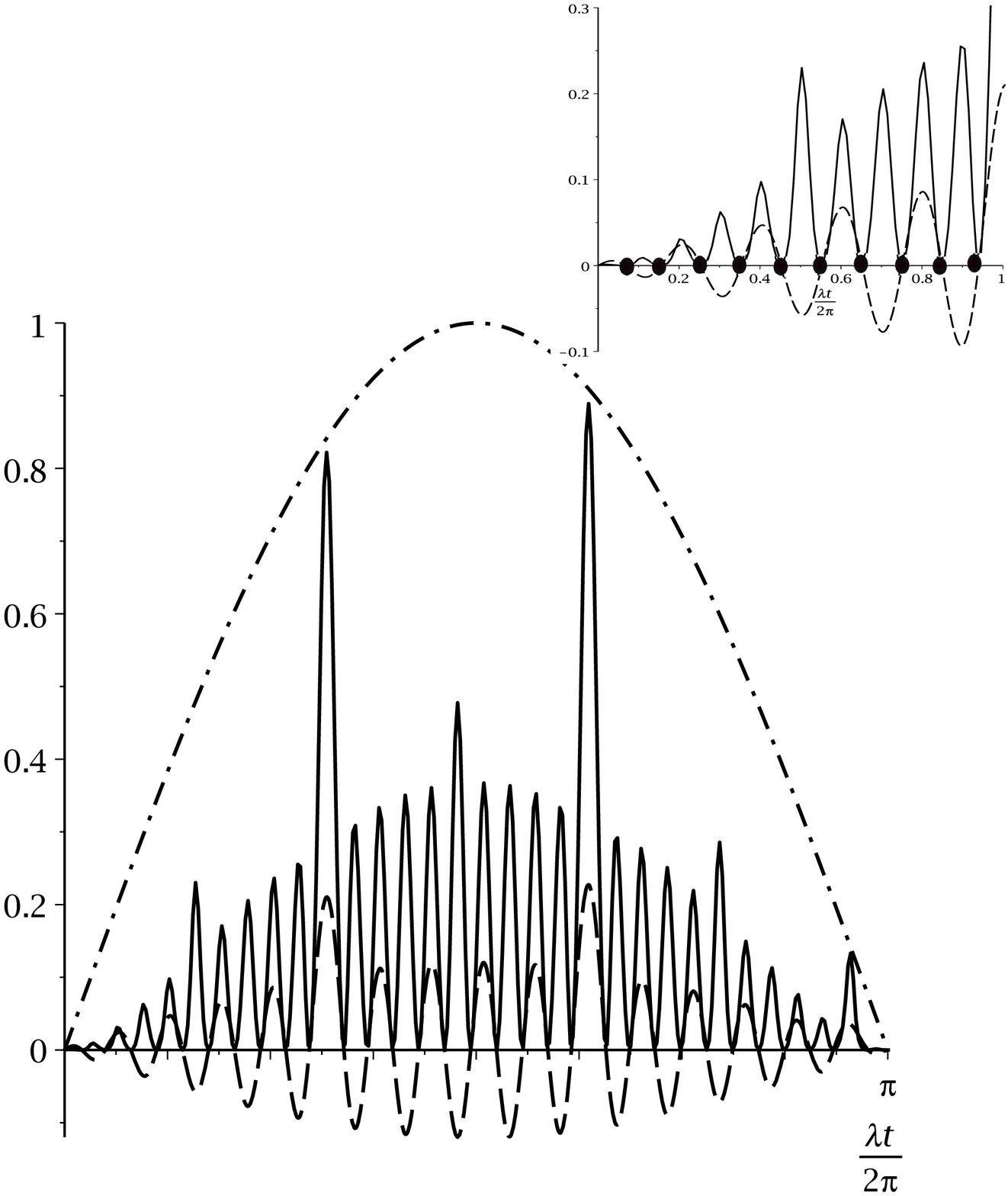}}
\subcaptionbox{\label{}}{\includegraphics[scale=0.31, angle=0.0, clip]{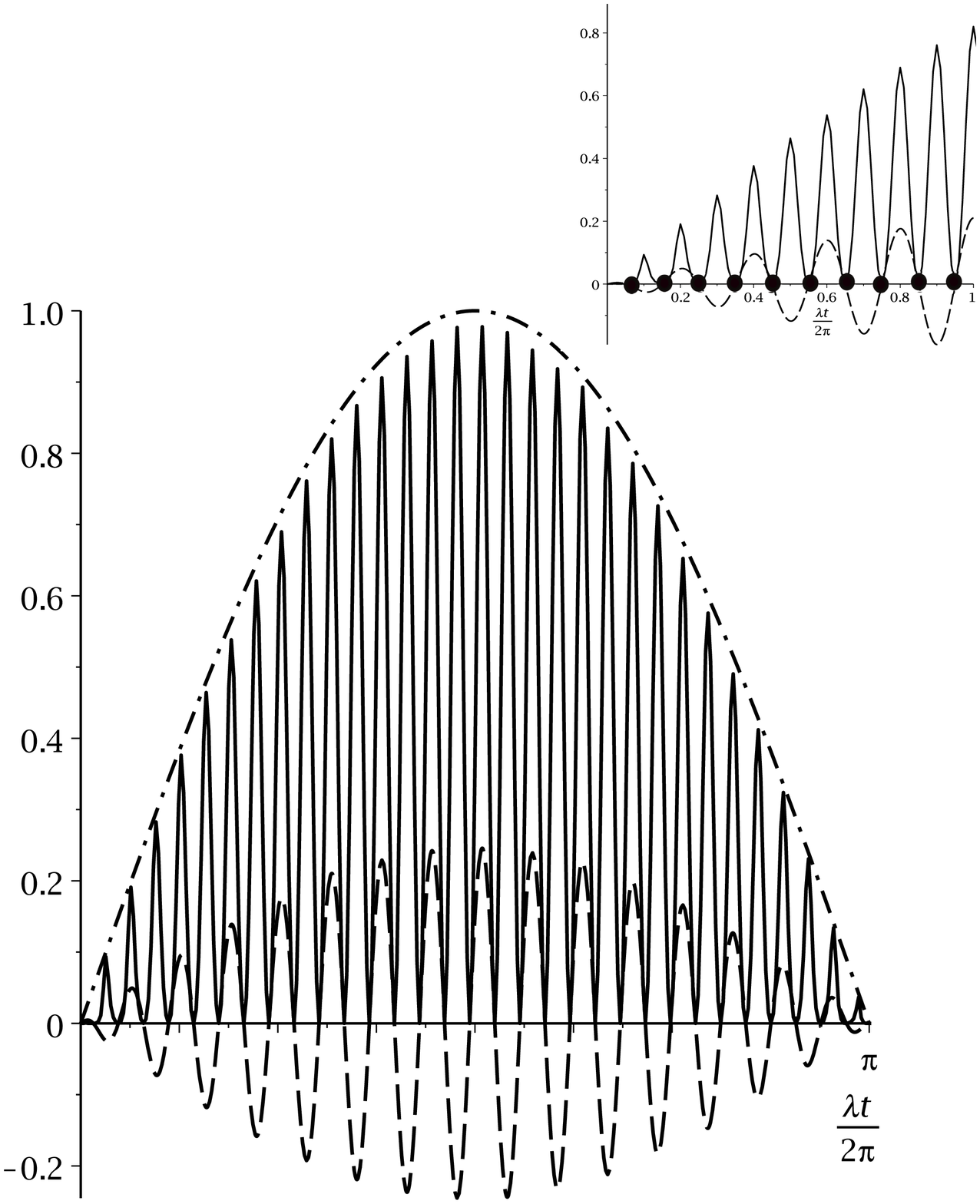}}
\caption{The connection between the Lee-Yang zeros of ten spins $1/2$ Ising ferromagnet with nearest-neighbor interaction (upper figures) and
concurrence of probe spins (solid curve), mean value of the probe spins operator $s_A^ys_B^z$
(\ref{connspinring}) (dashed curve) (lower figures). The results are presented for different temperatures: (a) $T=J^b/4$ and (b) $T=J^b/8$.
The upper figures show the positions of the Lee-Yang zeros of bath in the complex plane. The lower figures display the positions
of the Lee-Yang zeros in the time domain. In the insets, the moments of time when the Lee-Yang zeros are reached are marked by the solid circles.
The dash-dotted curve shows behavior of concurrence of the probe spins without the influence of the bath. It is well seen both in complex planes and in the time domain,
that the further decreasing of temperature leads to a uniform separation of zeros. The interaction between the probe spins is
isotropic and is defined by the coupling constant $J=\lambda/2\pi$.}
\label{lyzspinring2}
\end{figure}

We present our results for the bath consisting of ten spins $1/2$. Using the transfer-matrix method \cite{Kramers1941,Baxter1982,Strecka2015}
we express the partition function of such a model in the form
\begin{eqnarray}
Z\left(\beta,h\right)=z_+^N+z_-^N,
\label{partfuncspinring}
\end{eqnarray}
where
\begin{eqnarray}
z_{\pm}=e^{\frac{\beta J^b}{4}}\left[\cosh\left(\frac{\beta h}{2}\right)\pm \sqrt{\sinh^2\left(\frac{\beta h}{2}\right)+e^{-\beta J^b}}\right].\nonumber
\end{eqnarray}
Let us project the initial state of the probe spins on the $x$-axis along the opposite directions as follows
$\vert\psi(0)\rangle=\frac{1}{2}\left[\vert \frac{1}{2},\frac{1}{2}\rangle - \vert \frac{1}{2},-\frac{1}{2}\rangle + \vert -\frac{1}{2},\frac{1}{2}\rangle - \vert -\frac{1}{2},-\frac{1}{2}\rangle \right]$.
Then we obtain the following connection between the partition function of the bath and correlation functions of the probe system
\begin{eqnarray}
\langle s_A^xs_B^z\rangle = 0, \quad \langle s_A^ys_B^z\rangle=\frac{Z\left(\beta,-i\lambda t/\beta\right)}{Z\left(\beta,h=0\right)}\frac{1}{4}\sin(Jt).
\label{connspinring}
\end{eqnarray}
Note that these equations are valid for an arbitrary number $N$ of spins in the bath. Also, it is important to note, that the concurrence
of probe spins in the moments of time $t_n$, when the partition function $Z\left(\beta,-i\lambda t/\beta\right)$ vanishes, is equal to zero,
$C\left(\rho_{AB}(t_n)\right)=0$. In Figs.~\ref{lyzspinring1} and \ref{lyzspinring2} we show the time-dependence of correlation functions
(\ref{connspinring}) and concurrence of the probe spins in the cases of ten spins in bath and different temperatures.

\section{Discussions and conclusions \label{sec5}}

In papers \cite{zerospartfuncspin3,zerospartfuncspin4} the relation between the Lee-Yang zeros of the quantum Ising ferromagnet
and decoherence of the probe spin was obtained. Later using these results the Lee-Yang zeros were experimentally detected
on the trimethylphosphite molecule \cite{zerospartfuncspin1}. In recent paper \cite{kuzmak2019} we generalized this problem to the case of
an arbitrary high-spin quantum bath and proposed the new measured values of the probe spin, such as magnetization, susceptibility
and higher derivatives of magnetization. In the present paper we proposed to study the Lee-Yang zeros of an arbitrary quantum Ising ferromagnet via exploration
of the physical values of two probe interacting spins. So, we considered the arbitrary quantum ferromagnetic Ising bath (\ref{bathham})
which interacts with two probe spin-$1/2$ through Ising interaction (\ref{intham}). The interaction between probe spins is defined by anisotropic Heisenberg
model (\ref{probeham}). Using the fact that these three parts of the total Hamiltonian mutually commute we obtained the relation
between the partition function of the bath with effective complex magnetic field and correlation functions of probe spins
(\ref{averagevalue1})-(\ref{averagevalue4}), (\ref{averagevalue6}). Thus, the observation of time-dependent behavior of correlation functions of the probe spins allows
to detect the partition function zeros. The moments of time when these correlations functions vanish correspond to the Lee-Yang zeros of the bath.

Since the probe spins interact between themselves, they can be in the entangled states during evolution. The entanglement often occurs
in problems related to quantum information. We showed that it can be used for study the thermodynamic properties of a spin system.
So, we explored the influence of the bath system on the value of entanglement of probe spins. As a result, we obtained the explicit expression for the
values of the concurrence (measure of entanglement) of probe system (\ref{concomega1}), (\ref{concomega2}) at the moments of time when
the bath achieved the Lee-Yang zeros. These values of the concurrence of the probe system correspond to the Lee-Yang zeros of the bath with an effective
complex magnetic field. So, investigation of the entanglement of the probe system allows to detect the zeros of the partition function of the bath.

We applied the results obtained in the paper to the spin-$1/2$ ring which is described by the 1D ferromagnetic Ising model
with nearest-neighbor interaction as a bath and two probe spin-$1/2$ placed in the center of ring. The interaction between the probe spins is
described by the isotropic Heisenberg model. We projected the probe spins at the initial moment of time on the $x$-axis along the opposite directions.
In this case, at the moments of time, when the bath achieves the Lee-Yang zeros, the entanglement of the probe spins is equal to zero.
In Figs.~\ref{lyzspinring1} and \ref{lyzspinring2} we showed the location of the Lee-Yang zeros on a unit circle and marked the corresponding points
in the graphs which represent the time dependence of the correlation functions and concurrence of probe spins for different temperatures.

As we mentioned earlier, the bath does not influence the state of the probe spins if they evolve on the subspace spanned by the
$\vert \frac{1}{2},-\frac{1}{2}\rangle$, $\vert -\frac{1}{2},\frac{1}{2}\rangle$ vectors. This is because the interaction couplings of both spins
with the bath are same. Thus, in order to provide the influence of the bath on this evolution it is enough to make these couplings different.
Moreover, the magnitude of this influence depends on the value of difference between interaction couplings.
However, such interaction does not commute with the Hamiltonian of probe system (\ref{probeham}) except the case of $J_{xx}=0$.
Thus, to obtain the relation between the Lee-Yang zeros of the bath and measured values of the probe system the interaction between the probe spins
should be defined by the Ising model. In this case, the density matrix of the probe system contains four types of partition functions
which differ by the form of an effective complex magnetic fields: $-i\lambda_A t/\beta$, $-i\lambda_B t/\beta$, $-i(\lambda_A+\lambda_B) t/\beta$
and $-i(\lambda_A-\lambda_B) t/\beta$, respectively. Here $\lambda_{A(B)}$ is the value of interaction of $A(B)$ spin with the bath spins.
So, to detect the Lee-Yang zeros of the partition functions with $-i\lambda_A t/\beta$, $-i\lambda_B t/\beta$ magnetic fields, the correlation functions
$\langle s_A^xs_B^z\rangle$ and $\langle s_A^ys_B^z\rangle$, $\langle s_A^zs_B^x\rangle$ and $\langle s_A^zs_B^y\rangle$ should be measured. In the cases of partition functions with the
$-i(\lambda_A\pm\lambda_B) t/\beta$ magnetic fields, to detect the Lee-Yang zeros the correlation functions $\langle s_A^xs_B^x\rangle\mp \langle s_A^ys_B^y\rangle$ should be measured.
Finally, it is important to note that if the initial state of the probe spins belongs to the subspaces spanned by the
$\vert \frac{1}{2},-\frac{1}{2}\rangle$, $\vert -\frac{1}{2},\frac{1}{2}\rangle$ and $\vert \frac{1}{2},\frac{1}{2}\rangle$, $\vert -\frac{1}{2},-\frac{1}{2}\rangle$
vectors than the concurrence of this system is defined by expressions
$C\left(\rho_{AB}(t)\right)=2\left\vert a_{\frac{1}{2},\mp\frac{1}{2}}a_{-\frac{1}{2},\pm\frac{1}{2}} Z\left(\beta, -i(\lambda_A\mp\lambda_B)t/\beta\right)/Z\left(\beta,h=0\right)\right\vert$.
This expression is similar to expression obtained in the case of the Heisenberg interaction between probe spins (\ref{woottersAB2}). As we can see, here
the concurrence of the probe system is also proportional to the partition function of the bath with an effective complex magnetic field. So, the zeros of the
concurrence correspond to the Lee-Yang zeros of the bath.

\section{Acknowledgements}
The authors thank Dr. Andrij Rovenchak for useful comments. This work was supported by Project FF-83F (No.~0119U002203) from the Ministry of Education and Science of Ukraine.

\begin{appendices}
\section{Derivation the eigenvalues of matrix $R$ \label{appa}}
\setcounter{equation}{0}
\renewcommand{\theequation}{A\arabic{equation}}

In this appendix using Wootters definition of concurrence (\ref{wootters}) we obtain the eigenvalues of the matrix $R$ constructed by density
matrix (\ref{probeevolpf}) with $h=0$ and $t=t_n$. Then the density matrix $\rho_{AB}(t_n)$ and matrix $\tilde{\rho}_{AB}(t_n)$ have the form
\begin{eqnarray}
&&\rho_{AB}(t_n)=\left( \begin{array}{ccccc}
\left\vert a_{\frac{1}{2},-\frac{1}{2}}(t_n)\right\vert^2  & a_{\frac{1}{2},-\frac{1}{2}}(t_n)a_{-\frac{1}{2},\frac{1}{2}}^*(t_n) \\
a_{\frac{1}{2},-\frac{1}{2}}^*(t_n)a_{-\frac{1}{2},\frac{1}{2}}(t_n) & \left\vert a_{-\frac{1}{2},\frac{1}{2}}(t_n)\right\vert^2
\end{array}\right)\nonumber\\
&&\oplus\left( \begin{array}{ccccc}
\left\vert a_{\frac{1}{2},\frac{1}{2}}\right\vert^2 & a_{\frac{1}{2},\frac{1}{2}}(t_n)a_{-\frac{1}{2},-\frac{1}{2}}^*(t_n)\frac{Z\left(\beta, -i2\lambda t_n/\beta\right)}{Z\left(\beta, h=0\right)}\\
a_{\frac{1}{2},\frac{1}{2}}^*(t_n)a_{-\frac{1}{2},-\frac{1}{2}}(t_n)\frac{Z\left(\beta, -i2\lambda t_n/\beta\right)}{Z\left(\beta, h=0\right)} & \left\vert a_{-\frac{1}{2},-\frac{1}{2}}\right\vert^2
\end{array}\right)\nonumber\\
\label{rho}
\end{eqnarray}
and
\begin{eqnarray}
&&\tilde{\rho}_{AB}(t_n)=\left( \begin{array}{ccccc}
\left\vert a_{-\frac{1}{2},\frac{1}{2}}(t_n)\right\vert^2  & a_{\frac{1}{2},-\frac{1}{2}}(t_n)a_{-\frac{1}{2},\frac{1}{2}}^*(t_n) \\
a_{\frac{1}{2},-\frac{1}{2}}^*(t_n)a_{-\frac{1}{2},\frac{1}{2}}(t_n) & \left\vert a_{\frac{1}{2},-\frac{1}{2}}(t_n)\right\vert^2
\end{array}\right)\nonumber\\
&&\oplus \left( \begin{array}{ccccc}
\left\vert a_{-\frac{1}{2},-\frac{1}{2}}\right\vert^2 & a_{\frac{1}{2},\frac{1}{2}}(t_n)a_{-\frac{1}{2},-\frac{1}{2}}^*(t_n)\frac{Z\left(\beta, -i2\lambda t_n/\beta\right)}{Z\left(\beta, h=0\right)}\\
a_{\frac{1}{2},\frac{1}{2}}^*(t_n)a_{-\frac{1}{2},-\frac{1}{2}}(t_n)\frac{Z\left(\beta, -i2\lambda t_n/\beta\right)}{Z\left(\beta, h=0\right)} & \left\vert a_{\frac{1}{2},\frac{1}{2}}\right\vert^2
\end{array}\right),\nonumber\\
\label{tilderho}
\end{eqnarray}
respectively. The matrix $\rho_{AB}(t_n)\tilde{\rho}_{AB}(t_n)$ takes the following form
\begin{eqnarray}
&&\rho_{AB}(t_n)\tilde{\rho}_{AB}(t_n)\nonumber\\
&&{\scriptsize =\left( \begin{array}{ccccc}
2\left\vert a_{\frac{1}{2},-\frac{1}{2}}(t_n)\right\vert^2 \left\vert a_{-\frac{1}{2},\frac{1}{2}}(t_n)\right\vert^2 & 2\left\vert a_{\frac{1}{2},-\frac{1}{2}}(t_n)\right\vert^2 a_{\frac{1}{2},-\frac{1}{2}}(t_n) a_{-\frac{1}{2},\frac{1}{2}}^*(t_n) \\
2\left\vert a_{-\frac{1}{2},\frac{1}{2}}(t_n)\right\vert^2 a_{\frac{1}{2},-\frac{1}{2}}^*(t_n) a_{-\frac{1}{2},\frac{1}{2}}(t_n) & 2\left\vert a_{\frac{1}{2},-\frac{1}{2}}(t_n)\right\vert^2 \left\vert a_{-\frac{1}{2},\frac{1}{2}}(t_n)\right\vert^2
\end{array}\right)}\nonumber\\
&&{\scriptsize \oplus \left( \begin{array}{ccccc}
\left\vert a_{\frac{1}{2},\frac{1}{2}}\right\vert^2 \left\vert a_{-\frac{1}{2},-\frac{1}{2}}\right\vert^2 \left[1+\left(\frac{Z\left(\beta,-i2\lambda t_n/\beta\right)}{Z\left(\beta, h=0\right)}\right)^2\right] & 2\left\vert a_{\frac{1}{2},\frac{1}{2}}\right\vert^2 a_{\frac{1}{2},\frac{1}{2}}(t_n) a_{-\frac{1}{2},-\frac{1}{2}}^*(t_n) \frac{Z\left(\beta,-i2\lambda t_n/\beta\right)}{Z\left(\beta, h=0\right)} \\
2\left\vert a_{-\frac{1}{2},-\frac{1}{2}}\right\vert^2 a_{\frac{1}{2},\frac{1}{2}}^*(t_n) a_{-\frac{1}{2},-\frac{1}{2}}(t_n) \frac{Z\left(\beta,-i2\lambda t_n/\beta\right)}{Z\left(\beta, h=0\right)} & \left\vert a_{\frac{1}{2},\frac{1}{2}}\right\vert^2 \left\vert a_{-\frac{1}{2},-\frac{1}{2}}\right\vert^2\left[1+\left(\frac{Z\left(\beta,-i2\lambda t_n/\beta\right)}{Z\left(\beta, h=0\right)}\right)^2\right]
\end{array}\right).}\nonumber\\
\label{rhotilderho}
\end{eqnarray}
The eigenvalues $\omega_i^2$ of this matrix satisfy the following equations
\begin{eqnarray}
&&\omega^2\left(\omega^2-4\left\vert a_{\frac{1}{2},-\frac{1}{2}}(t_n)\right\vert^2 \left\vert a_{-\frac{1}{2},\frac{1}{2}}(t_n)\right\vert^2\right)=0,\nonumber\\
&&\omega^4-2\omega^2\left\vert a_{\frac{1}{2},\frac{1}{2}}\right\vert^2 \left\vert a_{-\frac{1}{2},-\frac{1}{2}}\right\vert^2 \left[1+\left(\frac{Z\left(\beta,-i2\lambda t_n/\beta\right)}{Z\left(\beta, h=0\right)}\right)^2\right]\nonumber\\
&&+\left\vert a_{\frac{1}{2},\frac{1}{2}}\right\vert^4 \left\vert a_{-\frac{1}{2},-\frac{1}{2}}\right\vert^4 \left[1-\left(\frac{Z\left(\beta,-i2\lambda t_n/\beta\right)}{Z\left(\beta, h=0\right)}\right)^2\right]^2=0.
\label{equtionforlambda}
\end{eqnarray}
Solving these equations we take into account only the positive solutions (\ref{woottersAB}).

\end{appendices}

\end{document}